\documentclass[aps,pre,twocolumn,superscriptaddress]{revtex4}
\usepackage{graphicx}
\usepackage{appendix}
\usepackage{amssymb}
\usepackage{amsmath}
\usepackage{color}
\usepackage[british]{babel}
\usepackage[normalem]{ulem}

\begin{document}

\title{Inverted critical adsorption of polyelectrolytes in confinement}
%planar slit, cylindrical pore, and spherical cavity}

\author{Sidney J. de Carvalho}
\affiliation{Institute of Biosciences, Letters and Exact Sciences, 
Sao Paulo State University, 15054-000 Sao Jose do Rio Preto, Brazil}
%\email{sidneyjc@ibilce.unesp.br}

\author{Ralf Metzler}
%\email{rmetzler@uni-potsdam.de}
\affiliation{Department of Physics, Tampere University of Technology, 33101 Tampere, Finland}
\affiliation{Institute for Physics \& Astronomy, University of Potsdam, 14476 Potsdam-Golm, Germany}

\author{Andrey G. Cherstvy}
\affiliation{Institute for Physics \& Astronomy, University of Potsdam, 14476 Potsdam-Golm, Germany}
%\email{a.cherstvy@gmail.com}

\date{\today}

\begin{abstract}

What are the fundamental laws for the adsorption of charged polymers onto 
oppositely charged surfaces, for convex, planar, and concave geometries? 
This question is at the heart of surface coating applications, 
various complex formation phenomena, as well 
as in the context of cellular and viral biophysics. 
It has been a long-standing challenge in theoretical polymer physics; 
for realistic systems the quantitative understanding 
is however often achievable only by computer simulations.
In this study, we present the findings of such extensive Monte-Carlo in silico experiments 
for polymer-surface adsorption in confined domains.
We study the inverted critical adsorption of finite-length polyelectrolytes 
in three fundamental geometries: planar slit, cylindrical pore, and spherical cavity. 
The scaling relations extracted from simulations for 
the critical surface charge density $\sigma_c$---defining the 
adsorption-desorption transition---are in excellent agreement 
with our analytical calculations based on the ground-state analysis of the Edwards equation. 
In particular, we confirm the magnitude
and scaling of $\sigma_c$ for the concave interfaces versus the Debye screening length 
$1/\kappa$ and the extent of confinement $a$ for these three interfaces for small $\kappa a$ values.
For large $\kappa a$ the critical adsorption condition approaches the planar limit.
The transition between the two regimes takes place when 
the radius of surface curvature or half of the slit thickness $a$ 
is of the order of $1/\kappa$. We also rationalize how $\sigma_c(\kappa)$ gets modified 
for semi-flexible versus  flexible  chains under external confinement. 
We examine the implications of the chain length onto critical 
adsorption---the effect often hard to tackle theoretically---putting 
an emphasis on polymers inside attractive spherical cavities. 
The applications of our findings to some biological systems are 
discussed, for instance the adsorption of nucleic acids onto the inner surfaces of 
cylindrical and spherical viral capsids. 

\end{abstract}

\maketitle

\section{Introduction} 

The adsorption of charged polymers or polyelectrolytes (PEs) 
%\footnotetext{Abbreviations: PE, polyelectrolyte; ES, electrostatic; WKB, Wentzel-Kramers-Brillouin;} 
onto oppositely charged surfaces 
\cite{korifei,netz03,dobr05,dobr-08,cw-aps-14} has a number of 
technological and biophysical applications including 
paper production \cite{paper,paper-2}, interface coating \cite{coating}, 
layer-by-layer formation \cite{polym,pems}, 
water desalination \cite{water,water-2}, and
stabilization of colloidal suspensions \cite{colloid,colloid-2}.
One distinguishes weak and strong PE-surface adsorption \cite{cw-aps-14}: 
weak PE-surface adsorption is governed by an interplay of 
energetic often electrostatic (ES) attraction of polyions onto an interface 
versus an entropic penalty accompanying the confinement 
\cite{degennes-confinement,holland} or the compression of the polymer chains 
near the surface \cite{wiegel-77,muthu87,muthu94}. 
Weak adsorption takes place for weakly charged PEs 
(partially neutralized by condensed counterions \cite{cc-manning,cc-grelet,cc-winkler}) and 
for moderately charged interfaces: such PE-surface binding is rather reversible. 
This contrasts an irreversible adsorption in the limit 
of strong PE-surface association \cite{cw-aps-14,helmut}.

\begin{figure}\includegraphics[width=6.5cm]{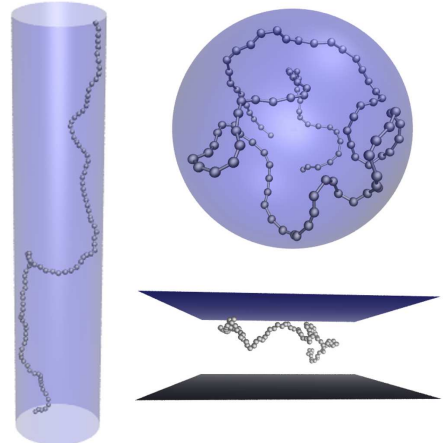}
\caption{Schematics of the inverted PE adsorption in confinement: 
planar slit, cylindrical pore, and spherical cavity. 
Video files illustrating the conformational 
changes of the polymer chain for conditions below and above the 
critical adsorption transition are 
presented in the Supplementary Material.}
\label{fig-scheme}\end{figure}

Weak PE adsorption onto surfaces of different geometries 
at varying conditions has been investigated  
in a number of recent theoretical \cite{cw-aps-14,cw-prl-06,muthu-pattern,belyi-viruses,
cw-jcp-06,cw-jpcb-07,cw-pccp-11,ac-biopol-12,netz99,netz00,sens99,
netz14-citations,muthu-confinement,krefeld-14,
forsman-es-nones,rudi-viruses,rudi-14-viruses}, experimental 
\cite{dubin-1,dubin-pore,dubin-2,dubin-3,dubin-06,dubin-4,dubin-11,
dubin-2-1-scaling,santore,bordi-12,pogi-pems,santer-ads}, and computer simulation
 \cite{stoll-1,stoll-03, stoll-05,stoll-02,china-11,
vattu-08,stoll-2,carillo,bachman-14-prl-cylinder-wrapping,shin-prx,
ccm-14-janus,sidney10,muthu11,sidney13, portugal-13,molina-14,sidney14,bachman-14} studies. 
PE adsorption onto patterned and corrugated surfaces 
\cite{muthu-pattern,krefeld-14,indians-patterned,gerda-14, 
bachman-14-pattern-prl,shakhnovich-pattern} as well as critical PE adsorption onto 
charged Janus net-neutral particles \cite{ccm-14-janus} was also examined. 
The properties of polymer adsorption inside cylindrical nanopores in porous glass were 
studied experimentally and by computer simulations in Ref. 
\cite{dubin-pore}. Possible effects of surface dielectric discontinuities on 
PE-surface adsorption were rationalized too  
\cite{cw-12-diel,stoll-09-diel,messina-diel}.

The critical adsorption describes the threshold conditions 
at which the ES-driven adsorption of PE chains first takes place in the system. 
This phase transition interrelates the condition for the interface surface charge 
density $\sigma$, the line charge density of the polymer $\rho$, 
the reciprocal Debye screening length in the solution $\kappa$, 
the ambient temperature $T$, and the polymer's Kuhn length $b$. The critical 
adsorption condition defines the relation between these important model 
parameters at the coexistence boundary of adsorbed versus desorbed chain
 conformations. Typically, a universal critical adsorption 
parameter can be constructed, \begin{equation} \delta_c=
\frac{24\pi a^3|\rho\sigma_c|}{\epsilon k_B T b},
\label{eq-delta-crit}\end{equation}  and its dependence on $\kappa$ 
governs the scaling of  $\sigma_c$---the critical surface charge density required for PE  
adsorption to take place---at varying salt conditions. 
Here $\epsilon$ is the dielectric constant of the medium, 
$a$ is the curvature radius of the adsorbing surface, and $k_B$ is the Boltzmann constant.
 For a long flexible nearly Gaussian charged polymers in front of 
a uniformly  oppositely charged plane the well-known result is \cite{wiegel-77} 
\begin{equation}\delta_c^{pl}(\kappa)\sim \sigma_c^{pl}(\kappa) \sim \kappa^3. \label{eq-crit-plane} 
\end{equation} The standard rationale  for increasing  $\sigma_c$ 
with the salt concentration $n_0$ is the requirement to compensate 
a stronger screening of ES attraction of the PE chain to 
the oppositely charged surface. For a symmetric 
1:1 electrolyte we have $\kappa^2=8 \pi l_B n_0$, 
where $l_B=e_0^2/(\epsilon k_B T)$ is the Bjerrum length. The peculiar cubical scaling of 
the critical charge density with $\kappa$ in Eq. (\ref{eq-delta-crit}) 
stems from the properties of the eigenfunctions 
of the corresponding Edwards equation for the 
conformations of a long polymer chain in the attractive 
Debye-H\"uckel potential  of the interface \cite{wiegel-77}.  In addition, some ES 
chain stiffening at low-salt conditions takes place impeding the PE-surface 
adsorption (see also Ref. \cite{forsman-es-nones} for non-ES effects in PE-surface adsorption).

For the convex cylindrical geometry (see Fig. \ref{fig-scheme}) a quadratic scaling is instead 
predicted by the Wentzel-Kramers-Brillouin (WKB) theory at low salt \cite{cw-pccp-11}, namely 
\begin{equation}\delta_c^{cyl}(\kappa a)\sim (\kappa a)^2, \label{eq-crit-cyl}\end{equation} 
while at high salinities and large rod radii $\kappa a \gg 1$ 
the planar limit (\ref{eq-crit-plane}) is recovered. For PE adsorption on the outside of 
oppositely charged spherical particles yields the linear 
dependence of $\sigma_c(\kappa a)$ in the limit $\kappa a \ll 1$
  \cite{cw-pccp-11}, \begin{equation} 
\delta_c^{sp}(\kappa a)\sim (\kappa a)^1.\label{eq-crit-sphere}\end{equation} Here $a$ is the 
radius of the cylinder or sphere. For more details on these 
scalings the reader is referred to Refs. 
\cite{cw-pccp-11,cw-aps-14,ccm-14-janus}. 

The systematic change in the $\sigma_c(\kappa a)$-scaling behavior from the 
planar interface via a cylinder to a sphere is in agreement with a number 
of experimental evidences from the Dubin's lab, see e.g.
 Ref. \cite{dubin-2-1-scaling} (and also the analysis in Ref. \cite{cw-jcp-06}). 
The experimental observations of critical 
PE adsorption are based on the complex formation of various 
polymers with oppositely charged particles and micelles of spherical 
and cylindrical geometry \cite{dubin-2-1-scaling}. 
These experimental findings indicate a weaker 
dependence of $\sigma_c$ on $\kappa a$ for more "convex" surfaces, 
as the adsorbing interfaces transfer from the planar 
to the cylindrical and finally to the spherical shape.

The adsorption transition of weak PEs under 
confinement \cite{muthu-confinement}---we call below the 
inverted critical adsorption---has a 
number of biologically relevant applications. 
Here, the term critical for confined PEs has 
the same meaning as for adsorption of charged polymers 
onto the planar and convex interfaces \cite{cw-aps-14}. 
For instance, the self-assembly of cylindrical and spherical single-stranded 
RNA viruses involves the adsorption of nucleic acids 
onto the inner virus capsid surface composed of protein building blocks  
\cite{belyi-viruses,muthu-viruses,rudi-viruses,twarok-14-viruses}. 
The capsid proteins are abundant in highly-basic 
flexible poly-peptide tails which trigger the adsorption of negatively-charged 
nucleic acids thus steering the self-assembly  \cite{hagan-13,hagan-14}. 
The known examples include nucleic acid encapsulation inside 
the cylindrical tobacco mosaic virus TMV  \cite{klug-tmv} and the icosahedral 
cowpea chlorotic mottle virus CCMV  \cite{ccmv,ccmv-2,belyi-viruses}.

For very long chains, the scaling relations for critical PE adsorption in inverted 
geometries were recently derived theoretically from the 
ground-state analysis of the Edwards equation for the Green function \cite{ac-biopol-12}.
The main subject of the current paper is the inverted weak adsorption of 
finite-length PE chains of varying stiffness in all three basic geometries, see Fig. \ref{fig-scheme}.
For long flexible polymers the critical adsorption conditions were 
obtained using the WKB method in Ref. \cite{cw-pccp-11}. Namely, 
Eqs. (5), (6), and (7) of Ref. \cite{cw-pccp-11} 
provide the dependence of $\sigma_c$ in the entire range of $\kappa a$. At low salt or 
strong confinement when $\kappa a \ll 1$ the critical 
adsorption parameter $\delta_c^{inv}$ for a planar 
slit was predicted to scale as \cite{ac-biopol-12} 
\begin{equation}\delta_c^{pl,inv}\sim 3C^2 (\kappa a)^1.
\label{eq-inv-slit}\end{equation}
Here and below the constant $C$ is of order unity, $C\approx 0.973$.
For long flexible PEs inside the oppositely charged 
cylinder this parameter reveals a plateau with a slowly varying logarithmic 
correction \cite{ac-biopol-12} 
\begin{equation}\delta_c^{cyl,inv}\sim \frac{3C^2}{0.116-\log(\kappa a)}.
\label{eq-inv-cyl}\end{equation} Finally, for a PE inside spherical cavities 
in the low salt limit the value of $\delta_c$ tends to 
saturate to a plateau  \cite{ac-biopol-12}
\begin{equation}\delta_c^{sp,inv}\sim 3C^2. \label{eq-inv-sp}\end{equation} 
These functional dependencies on $\kappa a$ are in stark 
contrast to the fast and monotonically increasing $\delta_c$ for the 
adsorption of PEs on the outside of cylindrical and spherical interfaces, 
Eqs. (\ref{eq-crit-cyl}) and (\ref{eq-crit-sphere}). In the opposite 
limit of loose confinement or high salt when $\kappa a \gg 1$ the theory predicts 
\begin{equation}\delta_c^{inv}=3C^2(\kappa a)^3/2\label{eq-3-2-large-salt}
\end{equation} for all three inverted geometries  
\cite{ac-biopol-12}. This latter limiting behavior was derived from the WKB approach in the 
limit of zero surface curvature \cite{cw-pccp-11,cw-aps-14}.
Note that for a finite-length polymer all the above 
mentioned standard ground-state based predictions for $\sigma_c$ need to be modified; the regular 
procedure is however not easy. This makes the findings of our computer 
simulations---the main focus of the current investigation---even 
more valuable for experimentally relevant situations. 

In the current paper we study by extensive Monte-Carlo simulations the 
properties of  critical PE adsorption in three basic inverted geometries, 
see Fig. \ref{fig-scheme}. We study the effects of the 
chain length, the polymer persistence, and systematically of 
the confinement size and solution salinity onto the critical 
surface charge density $\sigma_c$. In Sec. \ref{sec-model} we present the 
details of the simulation model and the data analysis algorithms. The main results 
on PE adsorption profiles and critical 
adsorption characteristics are described in Sec. \ref{sec-results}. 
We discuss the physical rationales behind the observed dependencies and 
the applications of our results in Sec. \ref{sec-discussion}.

\begin{figure}
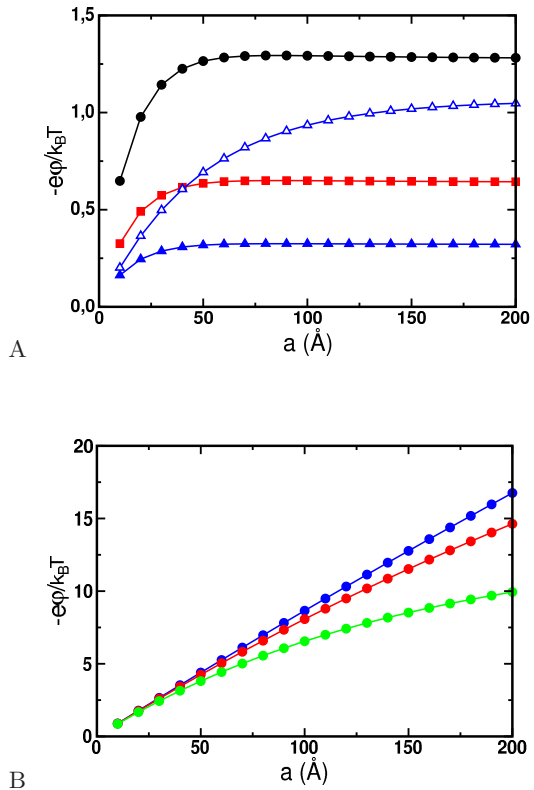

A~~\includegraphics[width=6.5cm]{fig2.eps}\vspace{1cm}
B~~\includegraphics[width=6.5cm]{fig2b.eps}
\caption{(A) Variation of the dimensionless 
ES potential $\Psi(a)=e_0\varphi(a)/(k_BT)$ at the boundary versus the radius of the spherical 
cavity plotted for: $\sigma=-0.1/(4 \pi)$ C/m$^2$ and $\kappa$=1/(30\AA)~ (filled circles), 
$\sigma=-0.1/(8 \pi)$ C/m$^2$ and $\kappa$=1/(30\AA)~ (filled squares), 
$\sigma=-0.1/(16 \pi)$ C/m$^2$ and $\kappa$=1/(30\AA)~ (filled triangles), 
$\sigma=-0.1/(16 \pi)$ C/m$^2$ and $\kappa$=1/(100\AA)~ (empty triangles). 
Note that C/m$^2 \approx e_0/$(16\AA$^2)$ that is $\approx6 \sigma_{\text{B-DNA}}$, 
where $\sigma_{\text{B-DNA}}$is the bare charge density of the B-DNA \cite{pccp-rev}. 
For large $a$ values the potential approaches the known planar result, 
$\varphi(a)=4 \pi \sigma/(\epsilon \kappa)$.
(B) The ES potential on the inner surface of spherical cavities
scales as $\Psi(a) \sim \kappa a$ for $\kappa a \ll 1$, 
whereas $\Psi(a) \to e_0 \varphi(a)/(k_B T)$ for $\kappa a \gg 1$. 
Parameters: $\sigma=-0.1/(4 \pi)$ C/m$^2$, 
$\kappa=1/(3000$\AA)~(blue symbols), $\kappa=1/(1000$\AA)~(red symbols), 
$\kappa=1/(300$\AA)~(green symbols).} \label{fig-es-pot}\end{figure}

\section{Model and Approximations}
\label{sec-model}

We here implement the same Metropolis Monte-Carlo simulation 
algorithm which has been successfully applied and tested by us recently for 
PE adsorption onto spherical \cite{sidney10}, cylindrical \cite{sidney13}, 
and spherical Janus \cite{ccm-14-janus} particles as well 
as for surface adsorption of pH-sensitive PEs  \cite{sidney14}. 
We refer the reader to Refs. \cite{ccm-14-janus,sidney10,sidney13} 
for more details on the simulation procedure. In brief, the polymer chain is modeled 
within the spring-bead model, with each monomer being 
a rigid sphere of radius $R_m=2$\AA~ carrying a point elementary 
charge $Z_m = e_0$ at its center. Neighboring beads are connected  
by the harmonic potential $$U_{str}(r) = K_r(r-r_0)^2/2,$$ with the elastic 
constant for bond stretching {$K_r = 1.0$N/m$^2$} and the 
inter-monomer equilibrium distance $r_0 = 7.0$\AA\ 
(as for single-stranded DNA \cite{pccp-rev}). The 
chain stiffness is given by the elastic potential 
 $$U_{el}(\theta) = K_{\theta}(\theta-\theta_0)^2/2,$$
where the force constant $K_{\theta}$ assumes the values such that the non-ES 
persistence length $l_{p,0}$ of the polymer ranges 
from {about} $8$ to $50$ \AA~ (a typical range for many real PEs). Here $\theta$ denotes the 
angle between the two successive bonds and $\theta_0 = \pi$. 
The mechanical persistence length for an uncharged chain $l_{p,0}$ was obtained in simulations via 
the relation $l_{p,0} = \langle R^2 \rangle^{1/2}/(1+\langle \cos \theta \rangle)$, where 
$\langle R^2 \rangle$ is the root-mean-squared monomer-monomer 
distance \cite{linse-lp}. The inter-chain excluded volume is accounted for by the 
standard hard-core repulsive potentials in simulations, 
as compared to the theoretical model \cite{ac-biopol-12}.

The repulsion of monomers at distance $r$ is 
given by the screened Coulomb potential
\begin{equation} U_{ES}(r)=\frac{Z_m^2 e^{-\kappa r}}{\epsilon r}.\end{equation}
% Here $\epsilon_s$ is the dielectric constant of the medium. 
The ES potential emerging in a slit with inter-plane distance  
$2a$, inside a cylinder or a sphere of radius $a$ 
were computed as the solutions of the linear Poisson-Boltzmann 
equation \cite{ac-biopol-12}. 
Below, we use the potentials denoted as $\Psi_{in,out}(r)$ in Ref. \cite{ac-biopol-12} to 
parametrize the strength of ES PE-surface attraction. For brevity, 
we do not provide the explicit analytical expressions here, 
instead showing the potential distributions in Fig. \ref{fig-es-pot}. 

The critical surface charge density $\sigma_c$ is defined in our simulations as the 
condition at which the PE binding energy to the interface exceeds the thermal energy,
\begin{equation}|E_b| \geq k_{B}T.\label{eq-ads-cond}\end{equation} 
Thus, even for the conditions when the polymer is not 
in direct contact with the surface but 
its total binding energy is lower than $E_b=-k_{B}T$, 
we consider the chain to be in the adsorbed state. 
{To compute the value of $\sigma_c$ for given values of the model parameters 
$\kappa$, $a$, and $N$, we perform the simulations for a set of surface charge densities and 
then determine the one for which the 
adsorption-desorption criterion (\ref{eq-ads-cond}) is satisfied.}
Here one can anticipate already that longer chains will require smaller surface charge densities 
$\sigma_c$ to be classified as adsorbed, as we indeed obtain 
from simulations, see below and also Ref. \cite{oshanesy-ads}. 
In the limit $a \to \infty$ the potentials  $\Psi_{in,out}(r)$ of Ref. \cite{ac-biopol-12} turn 
into $\Psi(r)$ for the corresponding isolated surfaces 
(see Fig. \ref{fig-es-pot}); the same holds for the properties of 
the adsorption-desorption transition, see below.  
 
There exists a number of differences between the inverted PE adsorption and the 
polymer-surface adsorption from a dilute, free-space solution. One feature is the 
presence of confining interfaces. They have different implications onto the polymer: for a 
planar slit the polymer is mobile in two 
dimensions, for a cylindrical tube the chain 
is free to move in one direction, and 
for a spherical cavity the polymer has no translational freedom at all. 
This progressively increasing confinement reduces the polymer 
conformational entropy \cite{shin-acs}, particularly upon adsorption on the interior 
of oppositely charged cylinders and spheres, see also Ref. \cite{mexico}.

We also note that in the low-salt limit the total PE persistence length, 
\begin{equation} l_p(\kappa) = l_{p,0}+l_{p}^{ES}(\kappa)=b/2,\label{eq-lp-total}\end{equation}
 acquires an ES component which is decreasing with the 
solution salinity. For flexible chains it obeys the 
scaling $l_p^{ES}(\kappa) \sim \kappa^{-1}$ \cite{dobr05} 
while for semi-flexible polymers  $l_p^{ES}(\kappa) 
\sim \kappa^{-2}$ \cite{netz03}. This fact is not accounted for in the theories of 
PE-surface adsorption \cite{cw-pccp-11,ac-biopol-12} 
yielding for $\delta_c^{inv}$ the 
scaling relations (\ref{eq-inv-slit}), (\ref{eq-inv-cyl}), 
(\ref{eq-inv-sp}).  This ES contribution  $l_p^{ES}$ should 
renormalize the scaling of $\sigma_c$ with $\kappa$ 
obtained from computer simulations at low salt, 
in accord with Eq. (\ref{eq-delta-crit}), 
see Ref. \cite{muthu87} the for planar and convex surfaces.
For concave adsorbing interfaces, such as the sphere's inner surface, 
due to this ES polymer stiffening the chains will tend to 
occupy regions of smaller curvature \cite{shin-acs}, 
as it is indeed observed upon "spooling" of double-stranded DNA inside 
bacteriophages \cite{marenduzzo-pnas}. 
The chains approach the interface because of bending energy 
minimization thus facilitating the ES-driven PE-surface binding, see below. 

An additional important parameter for confined 
PE-surface adsorption is the polymer's volume density. 
In the theory \cite{cw-aps-14,cw-pccp-11} the PE adsorption 
typically takes place from a very dilute polymer solution, which is not 
the case for confined inverted-adsorption situations, 
where the net polymer density is finite, see below. 

Let us now briefly discuss some approximations involved in the current study. 

a) We use the Debye-H\"uckel theory to compute the ES potentials 
near the interfaces and between the polymer monomers. 
This approach is valid for weakly charged 
systems and for an appreciable amount of salt in the solution $n_0$, 
when the ES potentials $|\Psi| \leq$25 mV (compare the panels of Fig. \ref{fig-es-pot}). 
The solution of the nonlinear Poisson-Boltzmann equation 
in curved geometries in the presence of salt is a formidable 
theoretical problem per se, often only solvable in some 
idealized limits (see Ref. \cite{lifson-pnas-cylinder} for a charged rod at $n_0=0$). 
Note that the linear ES theory often overestimates the 
magnitude of the potential emerging near highly charged 
interfaces (see Fig. 2 of Ref. \cite{agc-07} and Ref. \cite{fogolari-review-es}).
Also note that particular in low-salt solutions, 
the effects of counterion release from the surface---on 
the level beyond the standard Poisson-Boltzmann approach 
with the cation concentration obeying $n(r)=n_0 e^{-\Psi(r)}$---might 
become relevant for PE-surface adsorption.

b) The WKB scaling relations presented in the Introduction
stem from the ground-state analysis of infinitely long flexible 
Gaussian chains in front of surfaces with the Debye-H\"uckel ES potential. 
Both these idealizations will not hold upon variation of $n_0$  
in a broad range, as we study below. The investigation of implications of 
the non-linear nature of the ES potential near highly charged surfaces 
is the subject of a separate 
investigation \cite{sidney-prep}. Also, the impact of the 
mutual influence of adsorbing PE chains onto the ES potential of 
the interface (charge regulation) can non-trivially impact the critical 
adsorption conditions in terms of $\sigma_c(\kappa)$ scaling. 
Moreover, we consider below the adsorption of a single PE chain from a dilute solution; 
at realistic conditions however several chains might adsorb simultaneously. 
Their mutual salt-dependent ES repulsion 
along the surface will have an effect, for instance, 
on the overall surface coverage by PEs. 
The latter is often measured experimentally for PE-surface adsorption 
from bulk solutions with a finite polymer concentration.
All these effects are experimentally relevant and 
will be considered elsewhere \cite{sidney-prep}. 

c) We implement the $k_B T$-based adsorption criterion (\ref{eq-ads-cond}) 
to identify the PE adsorption-desorption threshold. 
This criterion is however somewhat arbitrary: 
for instance, one can classify the adsorption 
threshold using polymer distributions with a single peak versus the 
double-peaked profiles emerging between the two confining interfaces, 
see e.g. Fig. \ref{fig-pol-pdf-lp-0} below. 
We assume reversibility and ergodicity \cite{rm-pccp-14} 
for the process of PE adsorption at 
all conditions (no irreversible binding). 
This assumption might not be valid, particularly 
at low salt when the binding of even several PE  
monomers to an interface with large ES potential might 
overcome the thermal energy. Also, in this limit the relative 
accuracy in defining the adsorption-desorption boundary becomes important and 
even small fluctuations $|\delta\sigma_c|/|\sigma_c|$ that can realize  
in experiments might cause sizable effects. Finally, the adsorption of 
one fragment of the chain is assumed not to affect the ES potential acting to 
attract other parts of the polymer. This might be important 
for pH- and potential-responsive surfaces (not a part of this study \cite{sidney-prep}).
We work in the single-chain limit and thus do not 
study the PE adsorption isotherm---the amount of polymer adsorbed 
for a varying bulk polymer concentration. 

Despite these simplifications and assumptions, our computational results reveal excellent 
agreement with the theoretical predictions in a wide range of model parameters, see below.

\section{Results}
\label{sec-results}

\subsection{Polymer density distribution}

\begin{figure}
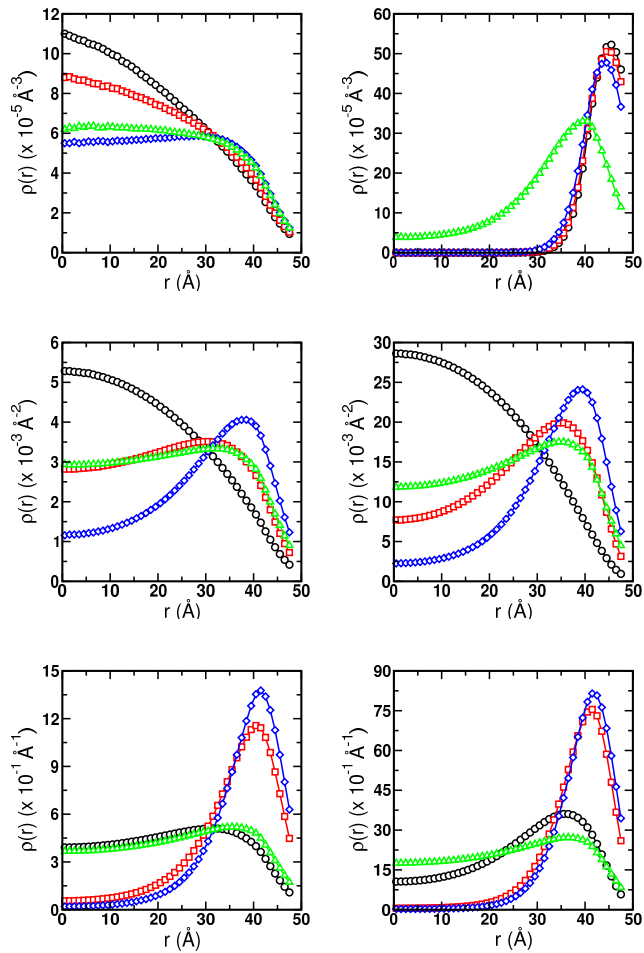

\centering
\includegraphics[width=4cm]{Distr_Sph_N20_kang0.eps}
\hspace{0.3cm}\vspace{0.5cm}
\includegraphics[width=4cm]{Distr_Sph_N100_kang0.eps}
\includegraphics[width=4cm]{Distr_Cyl_N20_kang0.eps}
\hspace{0.3cm}\vspace{0.5cm}
\includegraphics[width=4cm]{Distr_Cyl_N100_kang0.eps}
\includegraphics[width=4cm]{Distr_Plan_N20_kang0.eps}
\hspace{0.3cm}\vspace{0.5cm}
\includegraphics[width=4cm]{Distr_Plan_N100_kang0.eps}
\caption{Distribution of polymer monomers for 
inverted adsorption of flexible PEs ($\l_{p,0}=8$\AA) in a spherical cavity 
(top panels), cylindrical pore (middle panels) and 
planar slit (bottom panels) with the surface charge density of $\sigma=-0.1/(4\pi)$ C/m$^2$. 
The radius of the sphere and cylinder is $a=50$ \AA, and the 
slit thickness is $2a=100$ \AA. The distance $r$ 
denotes the separation from the centre of the 
confining space. The degree of chain 
polymerization is $N=20$ (left panels) and $N=100$ (right panels). 
The salt concentration $n_0$ is varied: $\kappa a = 0.1$ (black), 
$\kappa a = 0.5$ (red symbols), $\kappa a = 1$ 
(blue symbols) and $\kappa a = 5$ (green symbols).}
\label{fig-pol-pdf-lp-0}\end{figure}

\begin{figure}
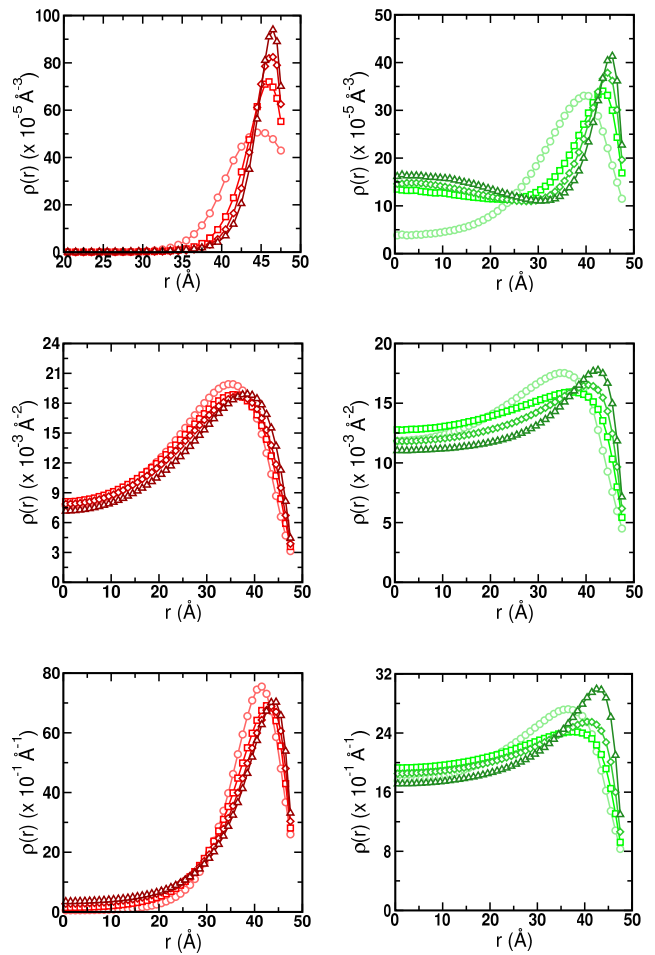

\includegraphics[width=4cm]{Distr_Sph_N100_ka_05.eps}
\hspace{0.3cm}\vspace{0.5cm}
\includegraphics[width=4cm]{Distr_Sph_N100_ka_5.eps}
\includegraphics[width=4cm]{Distr_Cyl_N100_ka_05.eps}
\hspace{0.3cm}\vspace{0.5cm}
\includegraphics[width=4cm]{Distr_Cyl_N100_ka_5.eps}
\includegraphics[width=4cm]{Distr_Plan_N100_ka_05.eps}
\hspace{0.3cm}\vspace{0.5cm}
\includegraphics[width=4cm]{Distr_Plan_N100_ka_5.eps}
\caption{Distribution of semi-flexible polymers inside a spherical cavity (top), 
cylindrical pore (middle) and planar slit (bottom panels) 
for the parameters of Fig. \ref{fig-pol-pdf-lp-0} and 
for $N=100$, $\kappa a = 0.5$ (left panels, reddish colors) 
and $\kappa a = 5.0$ (right panels, greenish colors). 
The color scheme matches that of Fig. \ref{fig-pol-pdf-lp-0}. 
The non-ES persistent length is $l_{p,0}=8$ (circles), 
23 (squares), 33 (diamonds) and 46\AA~(triangles).}
\label{fig-pol-pdf-lp}\end{figure}

First, we examine the distribution of polymer monomers, $\rho(r)$, in the 
three basic inverted geometries. For a fixed degree of the external confinement, 
the evolution of the $\rho(r)$ profiles with varying $n_0$ 
reflects the positioning of the adsorption-desorption boundary. 
We start with relatively flexible chains confined into a 
spherical cavity and a cylindrical pore; we use $\l_{p,0}=8$\AA~ 
for flexible chain results below. If the unperturbed 
radius of gyration of the chain exceeds the cavity 
dimensions, $\sqrt{\left<R_g^2\right>} \gtrsim a$, volume 
exclusion creates a force pushing the polymer  
towards the surface, see Fig. \ref{fig-pol-pdf-lp-0}.  This is 
particularly pronounced for spherical cavities, with strongly 
restricted chains. With increasing ionic strength up 
to $\kappa a = 1$ the monomer accumulation near the 
surface gets facilitated. From $\kappa a = 1$ to $\kappa a =5$ this behavior 
gets inverted, as at large $\kappa a$ the conditions 
are close to or above  the threshold of polymer desorption,  
and in addition the ES term in the chain 
persistence length gets smaller. This 
makes the polymer chains effectively more flexible, 
they are attracted weaker to the interface, and the polymers occupy 
the bulk of the cavity more readily. In this situation we thus find a single 
peak of the polymer distribution in the centre of the confining space, whereas 
for small $\kappa a $ the PE peak emerges near the 
oppositely charged surface in confined geometries.

The effect of the polymer's mechanical persistence onto PE  
distribution in confined spaces is illustrated in Fig. 
\ref{fig-pol-pdf-lp}. Progressively stiffer PE 
chains prefer to occupy the peripheral regions of the cavity due to a 
lower bending energy penalty for the chain arrangements with 
larger radii of curvature -- the effect particularly pronounced 
for spherical cavities, see the top row in Fig. \ref{fig-pol-pdf-lp}.
These latter trends are in line with the results of our 
recent simulations for polymer chains inside 
inert or non-attractive spherical cavities in the presence of 
macromolecular crowding \cite{shin-acs}, see also Refs. \cite{shin-njp,koreans-sm}. 
Moreover, the loss of configurational entropy for the arrangement of 
more persistent chains near the cavity surface is smaller as 
compared to the flexible ones. Fig. \ref{fig-pol-pdf-lp} 
shows that the deviations for semi-flexible versus flexible 
chains become smaller as we go from PE adsorption inside a spherical cavity to PE adsorption 
inside a cylindrical pore and finally to PE adsorption inside a planar slit 
(respectively, the top, middle and bottom panels of Fig. \ref{fig-pol-pdf-lp}). The physical 
reason is again the the number of the polymer's degrees of freedom available 
in the corresponding geometries. For instance, for a planar slit the bending 
energy of semi-flexible chains has nearly no implications on 
the amount of the polymer near the adsorbing interface, in stark contrast to the 
spherical cavity for which a severe chain bending is unavoidable \cite{shin-acs}, 
compare the panels in  Fig. \ref{fig-pol-pdf-lp}. As we demonstrate below, this 
polymer bending energy in spherical and 
cylindrical confinement has non-trivial effects onto the 
critical surface charge density $\sigma_c(\kappa a)$ of the adsorption-desorption transition. 

%\subsection{Amount of polymer adsorbed}

\begin{figure}[t!]
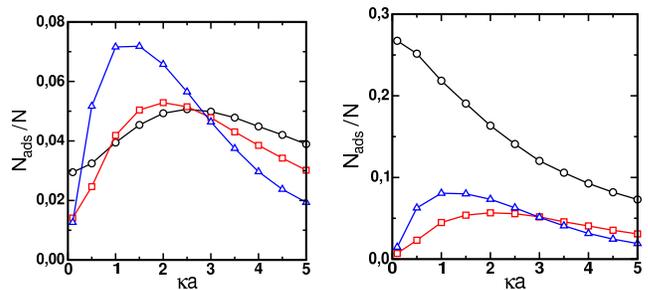

\includegraphics[width=4cm]{fig5-20.eps}\hspace{0.3cm}
\includegraphics[width=4cm]{fig5-100.eps}
\caption{Fraction of polymer monomers within 4\AA~from the adsorbing surface 
for inverted PE adsorption into a spherical cavity (black), 
cylindrical pore (red) and planar slit 
(blue symbols) with $\sigma=-0.1/(4\pi)$C/m$^2$ 
(above the critical adsorption transition), plotted for varying solution salinity. 
The radius of spherical cavity and the cylindrical tube is $a=50$ \AA~and 
the inter-plane distance for the slit is $2a=100$ \AA;  
$N=20$ (left panel) and $N=100$ (right panel); $\l_{p,0}=8$\AA.}
\label{fig-ads-fraction}\end{figure}

A relevant experimental question for super-critical PE adsorption 
is the amount of the polymer adsorbed on the surface. 
In Fig. \ref{fig-ads-fraction} we quantify how this 
amount changes with $\kappa a$ for the three adsorption geometries 
for the single-chain adsorption simulated. 
Namely, for $\sigma > \sigma_c$ we analyzed the 
PE profiles formed near the interfaces for the inverted 
polymer adsorption. We evaluate the fraction of the polymer chain in the region 
close to the adsorbing interface, $N_{ads}/N$. This fraction is a stationary quantity;  
we do not consider here the kinetics of PE adsorption 
for $\sigma > \sigma_c$ \cite{dobr05,stuart-kinetics,granick-kinetics}.
We illustrate the behavior of this fraction in Fig. \ref{fig-ads-fraction} 
versus the reciprocal Debye screening length, $\kappa$. We find that the 
amount of PE adsorbed within this first  layer near the 
interface is often a non-monotonic function of the salt concentration  $n_0$.
 One physical reason for this is a shorter ES persistence length of PEs 
and weaker polymer-surface charge-mediated binding 
as  $n_0$ increases. This non-monotonicity can
 be anticipated already from the evolution of PE profiles in Fig. 
\ref{fig-pol-pdf-lp-0} in the proximity of the adsorbing 
surface as $\kappa$ increases. Here, we refer the 
reader to the studies in Refs. \cite{santore,forsman-es-nones} for 
experimental evidence and theoretical predictions of non-monotonic effects of added salt 
on the amount of adsorbed PE chains. In realistic multi-chain systems, 
higher solution salinities effect softer PE chains, weaker PE-surface ES attraction, but 
also a weaker ES repulsion between the already adsorbed polymer coils. The 
interplay of these effects might yield a non-monotonic behavior 
of the mass-per-area of adsorbed PEs with varying $\kappa$.

The width of the PE profile $w$ together with the mass 
of adsorbed PE per area are the experimentally relevant quantities. 
As we work in the single chain limit the latter will not be considered. 
For inverted PE adsorption, the width of the adsorbed layer is 
 expected to be a non-monotonic function of $\kappa$ too. 
Note however that the standard definition of $w$ implemented for instance for the 
adsorption of an isolated PE chain onto an attractive 
surface \cite{cw-aps-14}---as the width 
of the polymer probability distribution function at its half-height---cannot 
be directly used for the current problem of inverted adsorption.

\subsection{Critical adsorption conditions }

\begin{figure}
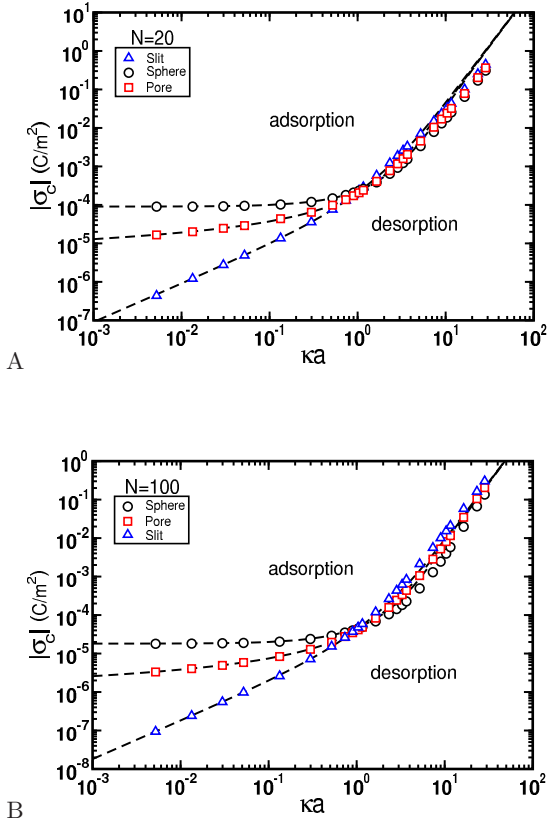

A\includegraphics[width=7cm]{fig6a.eps}\vspace{1cm}
B\includegraphics[width=7cm]{fig6b.eps}
\caption{Critical surface charge density $\sigma_c$ for inverted 
adsorption of flexible PEs ($\l_{p,0}=8$\AA) into a planar slit (blue symbols), 
cylindrical tube (red symbols), and spherical cavity (black symbols), 
plotted for varying solution salinity. 
The dashed lines are the full theoretical asymptotes for $\delta_c$
\cite{ac-biopol-12}, plotted for $\sigma_c$ using Eq. (\ref{eq-sigma-magnitude}). 
The corresponding low-salt limits are given by Eqs. (\ref{eq-inv-slit}), 
(\ref{eq-inv-cyl}), (\ref{eq-inv-sp}); the high-salt 
or the planar limit is Eq. (\ref{eq-3-2-large-salt}). 
Note the inverse positioning of the curves in the region of $\kappa a \gg 1$. Parameters:  
$a=50$ \AA~(1/2 of the slit thickness, the cylinder and 
sphere radii), the polymerization degree is {$N=20$} (panel A) and 
  {$N=100$} (panel B). On a standard 3-3.5 GHz workstation every curve on these graphs 
requires some 180 h and 900 h of computational time for 
chains of $N = 20$ and $N = 100$ monomers, respectively.}
\label{fig-crit-ads-all}\end{figure}

Now we turn to the main objective of the current study,
 the scaling of the critical surface charge density at the adsorption-desorption transition.
For the inverted PE adsorption in confined geometries, 
we find that $\sigma_c$ varies with $\kappa a$ as shown in Fig. \ref{fig-crit-ads-all}, 
revealing an excellent agreement with the theory developed in Ref. \cite{ac-biopol-12}.
 As expected, the scaling of the critical surface 
charge density is very different in the limit of small and large $\kappa a $ 
values, as prescribed by Eqs. (\ref{eq-inv-slit}), (\ref{eq-inv-cyl}), 
(\ref{eq-inv-sp}) and Eq. (\ref{eq-3-2-large-salt}), respectively. 
In the low-salt limit spherical cavities necessitate larger surface charge densities than the 
cylindrical tubes; the latter in turn need larger surface charges than 
the planar slits in order to reach the 
same degree of polymer binding (\ref{eq-ads-cond}), 
compare the curves in Fig. \ref{fig-crit-ads-all}. 
We attribute this reduction of $\sigma_c^{inv}$ to a progressively smaller 
penalty of entropic confinement of flexible PE chains.
Remarkably, even the reversed order of the 
critical adsorption curves in the limit of large $\kappa a$, 
as compared to the low-salt limit, is precisely reproduced in our simulations, 
in accord with the theory \cite{ac-biopol-12}. This reversed order at high salt is particularly
 well pronounced for longer chains, Fig. \ref{fig-crit-ads-all}B.  
The change in the $\sigma_c(\kappa a)$ scaling behavior, from the low-salt prediction to the 
high-salt asymptote (\ref{eq-3-2-large-salt}), occurs at $\kappa a \sim 1$, 
for any chain length, compare the panels in Fig. \ref{fig-crit-ads-all}. 
The effects of the surface curvature on the adsorption-desorption transition point 
is thus universal and start to be important for the conditions of low-salt 
and large surface curvature when $\kappa \lesssim 1/a$. 
Fig. \ref{fig-crit-ads-all} is the central result of the current study.

For progressively longer PE chains the values of $\sigma_c$ decrease, preserving however 
the overall scaling relations in the limit of low salt and the approach to 
the universal asymptote at high salt, 
compare Figs. \ref{fig-crit-ads-all}A and \ref{fig-crit-ads-all}B. 
According to our adsorption criterion, Eq.  (\ref{eq-ads-cond}), 
longer chains will clearly accumulate the same binding energy for 
smaller $|\sigma|$. A similar behavior was observed for 
PE adsorption onto Janus particles \cite{ccm-14-janus}: 
the entire $\sigma_c(\kappa a)$ dependence shifts down for longer chains but 
preserves the scaling properties for small and large $\kappa a$ values. 

In Fig. \ref{fig-crit-ads-all} flexible chains are considered, with 
the non-ES persistence length of $l_{p,0} \lesssim 10$ \AA. 
For a spherical cavity, the theoretical results of 
Ref. \cite{ac-biopol-12} for $\sigma_c$ give an excellent agreement 
with the results of our simulations for $N=100$ chains if we set 
$b=2 l_{p,0} \approx 4$ \AA, a realistic value for flexible chains simulated. 
This follows from Eq. (\ref{eq-inv-sp}) for PEs inside spherical 
cavities for $\kappa a \ll 1$  from which the critical 
charge density can be recalculated in units C/m$^2$ as 
\begin{equation}\frac{\sigma_c^{sp,inv}}{\text{C}/\text{m}^2}=
3 C^2 \frac{16 b r_0}{24 \pi a^3 l_B} \approx  
1.9 \times 10^{-5}.\label{eq-sigma-magnitude}\end{equation} 
The same Kuhn length $b$ %, found above for PE-inside-sphere adsorption, 
is then used to compute the full $\sigma_c^{inv}(\kappa)$ asymptotes 
%for the cylindrical and planar inverted adsorption
from the theoretically predicted \cite{ac-biopol-12} 
relations for $\delta_c^{cyl,inv}(\kappa)$ and $\delta_c^{pl,inv}(\kappa)$, 
according to Eq. (\ref{eq-delta-crit}).

\begin{figure}[t!]%\vspace{1cm}
\includegraphics[width=7cm]{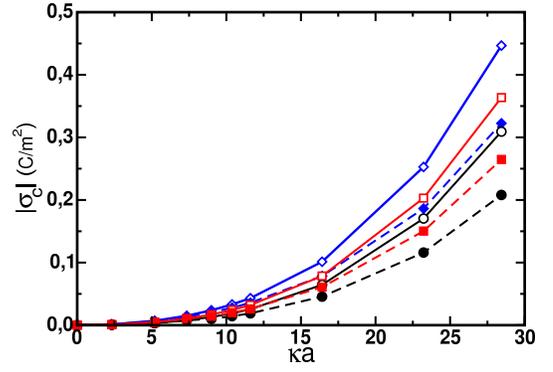}
\caption{The same as in Fig. \ref{fig-crit-ads-all} for 
inverted adsorption inside a spherical cavity (black),  
cylindrical tube (red), and planar slit (blue symbols) 
for varying $\kappa$. The results are plotted in the linear scale. 
The polymer stiffness is $l_{p,0}=8$ \AA~ (open symbols) and 
$l_{p,0}=50$\AA~(full symbols). Parameters:  $a=54$ \AA~ and $N=20$.}
\label{fig-crit-stiff}\end{figure}

\begin{figure}[t!]\vspace{1cm}
\includegraphics[width=6.5cm]{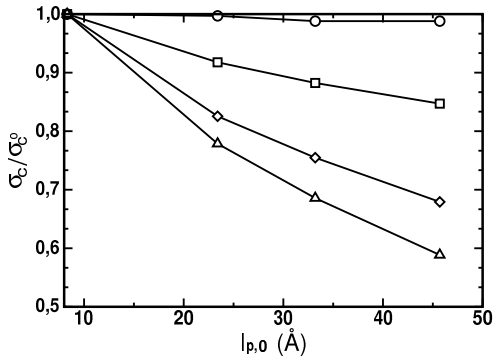}
\caption{Critical adsorption charge density for persistent 
versus flexible polymers  inside a 
spherical cavity, $\sigma_c/\sigma_c^0$, computed for varying non-ES polymer persistence length. 
Parameters: $a=50$ \AA, $N=100$, $\kappa a = 1$ (circles), $5$ (squares),
$8$ (diamonds) and $10$ (triangles).}
%The values of $l_{p,0}$ are the same as in Fig. \ref{fig-pol-pdf-lp}.}
\label{fig-crit-ads-all-persistent}\end{figure}

\begin{figure}\includegraphics[width=7cm]{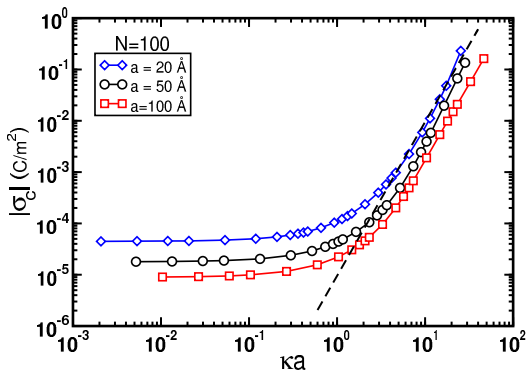}%\vspace{1cm}
\caption{The same as in Fig. \ref{fig-crit-ads-all} but 
for varying radius $a$ of the spherical cavities. 
The high-salt asymptote (\ref{eq-3-2-large-salt}) 
for the charge density (in units of C/m$^2$) is obtained from Eq. (\ref{eq-3-2-large-salt}) as
$\sigma_c(\kappa)/[\text{C}/\text{m}^2]=
%\frac{3}{2} C^2 (\kappa a)^3 \frac{16 b r_0}{24 \pi a^3 l_B}=
C^2 \kappa^3 b r_0/(\pi l_B )\sim \kappa^3$, the dashed 
line. Parameters: $N=100$, $a=50$\AA, $\l_{p,0}=8$\AA.} \label{fig-crit-ads-a}\end{figure}

\begin{figure}%\vspace{1cm}
\includegraphics[width=7cm]{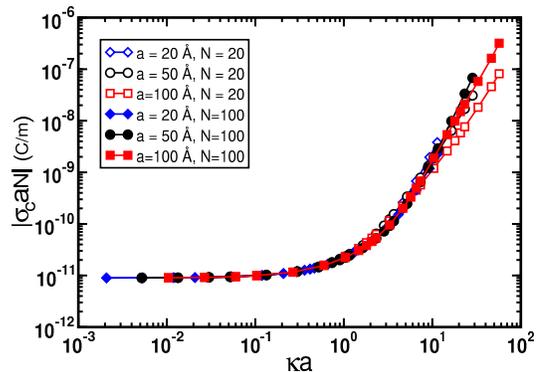}
\caption{Universal rescaled surface charge density for critical 
PE adsorption inside spherical cavities of varying radii and for 
polymers of different lengths, plotted for varying $\kappa$; $l_{p,0}=8$\AA.}
\label{fig-crit-ads-all-a-variation}\end{figure}

Due to the adsorption criterion implemented, for $N=20$ chains $\sigma_c$ 
is nearly 100/20=5 times larger than for $N=100$ polymers, 
compare the panels in Fig. \ref{fig-crit-ads-all} and see also 
the universal curves in Fig. \ref{fig-crit-ads-all-a-variation}.
Also note that for $\kappa a \gg 1$ our simulations of the PE adsorption 
under confinement yield the $\sigma_c(\kappa) \sim \kappa^3$ scaling behavior, as anticipated 
for polymers with a salinity-independent Kuhn length $b$ \cite{wiegel-77}. 
This is in contrast for instance to PE adsorption on the outside of spherical particles, 
where the effects of ES persistence are important
and our simulations in the high-salt limit give 
$\sigma_c(\kappa) \sim \kappa^{1.9}$ scaling instead, 
see Fig. 8A in Ref. \cite{ccm-14-janus}. 

We also examined the dependence of the critical adsorption 
conditions for more persistent chains in all three adsorption geometries, 
see Figs. \ref{fig-crit-stiff} and \ref{fig-crit-ads-all-persistent}. 
Fig. \ref{fig-crit-stiff} shows that for more persistent chains the magnitude of 
$\sigma_c$ decreases for the adsorption onto 
the planar slit, inside cylindrical pore, and spherical cavities.
We find that particularly for PE adsorption 
inside spherical cavities the magnitude $|\sigma_c|$ decreases due to a bending-energy 
driven localization of polymers near the cavity surface, see Fig.
 \ref{fig-crit-ads-all-persistent}. For PE-sphere inverted adsorption, 
stiffer PEs prefer to stay closer to the adsorbing interface thus 
reducing the value of $\sigma_c$. The precise 
behavior of $\sigma_c$ as a function of $\kappa a $ shows that 
the deviations from the flexible chain results become progressively 
larger for more persistent chains and at larger $\kappa a $ values, 
see Fig. \ref{fig-crit-ads-all-persistent}. 
The latter is not surprising because at high-salt conditions 
the ES contribution to the polymer persistence gets reduced and 
the PE stiffness is dominated by its 
mechanical part $l_{p,0}$, see Eq. (\ref{eq-lp-total}).

The question arises whether for inverted PE critical 
adsorption the variation of the confinement degree or salt concentration 
gives rise to different $\sigma_c$ behaviors? We showed that for adsorption of finite-length PEs 
onto spherical Janus particles \cite{ccm-14-janus} there was no universal parameter $\kappa a$
that would combine the curvature and salinity 
effects on the critical adsorption properties $\sigma_c(\kappa a)$.
The inverted critical PE adsorption is also quite different if one 
varies the size of the confined cavities, tubes, and slits or the solution salinity, 
 compare the curves in Fig. \ref{fig-crit-ads-a} for spherical confinement.
Longer chains require smaller surface charge densities to get adsorbed
 and for smaller sizes of spherical cavities the value of $\sigma_c$ increases. 
This can be understood from the variation of the ES surface potential in spherical 
cavities presented in Fig. \ref{fig-es-pot}b showing that 
$\Psi(a) \approx 2 (\kappa a) e_0 \varphi(a)/(k_B T)
 \sim \kappa a$ for $\kappa a \ll 1$ \cite{ac-biopol-12}. 
For inverted critical PE adsorption we obtain that indeed there exists no 
universal parameter $\kappa a $. This at first sight disagrees with 
the theoretical results of Ref. \cite{ac-biopol-12}. 
For the finite-length PEs with varying ES persistence studied in 
our simulations this disagreement is however not surprising, as compared to 
infinitely long flexible salt-insensitive polymers studied in the 
theoretical idealization \cite{cw-aps-14,ac-biopol-12}. 

Fig. \ref{fig-crit-ads-all-a-variation} illustrates the behavior of the 
rescaled critical surface charge density for inverted PE-sphere adsorption, 
$\sigma_c a N$. This combination accounts for the peculiar features of the 
variation of the ES potential with the cavity radius (Fig. \ref{fig-es-pot}b) 
and the adsorption condition used in the simulations, Eq. (\ref{eq-ads-cond}). 
We find a universal collapse of this renormalized parameter for the critical 
adsorption curves for different chain lengths and cavity sizes, for 
a wide range of variation of the solution salinity. 

\section{Discussion}% and Outlook}
\label{sec-discussion}

In this study, we have employed extensive Monte-Carlo computer 
simulations to unveil the physical properties of PE adsorption in confined 
spaces, considering polymer chains inside a planar 
slit, a cylindrical pore, and a spherical cavity.  We rationalized 
the position of the adsorption-desorption transition upon variation 
of various physical parameters such as the extent of the external confinement, the 
salinity of the solution, the chain length, and the bare persistence length of the polymer. 
We have demonstrated how the well-known cubic scaling of the critical 
surface charge density with the reciprocal Debye screening length $\kappa$ 
gets non-trivially modified. Namely, in the limit of $\kappa a \ll 1$---small solution 
salinities or large surface curvature $1/a$---for 
the critical adsorption condition for PEs under confinement 
splits for the three fundamental geometries. 
We illustrate this behavior in Fig. \ref{fig-crit-ads-all} 
that is the main result of this study. 
Our results revealed a remarkable quantitative agreement with the recent 
theoretical predictions for the same systems \cite{ac-biopol-12}. 
The simulation approach enabled us to vary the polymer length 
and PE persistence, which are often quite problematic 
to be properly implemented from the first theoretical principles \cite{muthu87}.
Also, we showed that for the critical adsorption onto concave surfaces more persistent chains require 
 smaller surface charge densities to get adsorbed.
For critical adsorption of PEs of varying polymerization degree 
$N$ inside spherical cavities of radius $a$, 
we found that the simulation results collapse onto a universal curve if 
the rescaled surface charge density is considered, 
namely $\sigma_c \to \sigma_c a N$.

Finally, only the static properties of PEs under confinement 
were considered in the current paper. 
It would be instructive as a next step to study the dynamics of 
charged polyions inside oppositely charged domains and cavities. 
In particular, the implications of polymer charge and adsorption to
 the spherical cavity interior can enrich the trends observed for 
looping kinetics of spherically-confined flexible and semi-flexible chains \cite{shin-acs}.
The video files of the Supplementary Material demonstrate, 
for instance, that the dynamics of PE chains in the 
adsorbed state is slowed down dramatically, as compared to desorbed configurations.
This surface-mediated polymer confinement is consistent 
with the ultraslow relaxation of confined DNA molecules detected in 
single-molecule experiments during viral packaging \cite{ultraslow}.
Future developments of the model will include the study of 
PE adsorption onto pH-responsive functionalized \cite{lasch-smart} curved surfaces, 
the implications of a nonlinear ES potential distribution on the position of the 
adsorption-desorption boundary $\sigma_c(\kappa,N)$ \cite{sidney-prep}, 
and the adsorption of polymer chains with heterogeneous charge distribution. 
The latter can be applied, for instance, to the surface-mediated adsorption of poly-peptide chains of 
partially folded proteins.

Let us discuss some possible applications of our findings. 
Polymer encapsidation inside oppositely charged cavities 
\cite{muthu11,rudi-14-viruses,hagan10-viruses,pccp-rev,bachman-14} is the 
fundamental mechanism of assembly of cylindrical 
and spherical single-stranded RNA viruses \cite{hagan-13,hagan-14}. 
This process employs a delicately tuned adsorption of negatively charged RNA chains onto 
 the positively charged interior of viral protein shells. 
Direct applications of our observations to the properties of real RNA viruses 
might however require the secondary RNA looped structure to be taken into account. 
The latter often pays an important role in viral assembly and the RNA packaging process 
\cite{rudi-14-viruses,schoot-secondary}. Branching and self-association in 
the structure of compacted RNA yield, for instance, a weak overcharging 
of the entire virion: on average the negative 
charge of the enclosed nucleic acid chain is about 1.6 times larger 
than the positive charge of the enveloping protein 
shell \cite{belyi-viruses}.

Another domain of possible applications includes the behavior of long DNA 
molecules in micro-fluidic devices involving nano-channels 
\cite{tapio-DNA-nano-channels,austin-DNA} with attractive walls. 
Having in mind some applications to cylindrical channels of non-trivial cross-section 
\cite{tapio-DNA-nano-channels-tirangular}, one can consider in the future 
the PE adsorption on the interior of tubes with more complicated geometries, 
e.g. rectangular or triangular rather than circular cylindrical channels.
Some applications of our findings to the description of charge effects of 
PE and DNA translocation though natural and 
synthetic nanopores \cite{paljulin} are also possible. 
One more immediate application of our results includes the problems of protein 
adsorption---both in their native form and in the denatured state---in various porous media. For instance, polymer dynamics and adsorption in 
sticky nano-channels of porous silicon studied in Ref. \cite{patrick} can pave the way for the 
selective separation of proteins from unknown mixtures, based on their 
surface charge and surface-adsorption properties.
Moreover, defining the critical adsorption conditions is of vital importance for the 
fabrication of responsive and permeable multilayer capsids. 
They are being formed via the alternating adsorption of oppositely charged PEs \cite{pems} 
and used for diagnostic and therapeutic purposes \cite{pems-2}. 
Finally, nano-structured polymer-functionalized porous materials are used in  
electro-chemical super-capacitors \cite{supercap-1,supercap-2} and 
our results on polymers under extreme confinement might 
find some future applications in this area as well.

\section{Acknowledgements}

Computer resources were supplied by the 
Center for Scientific Computing (NCC/GridUNESP) of 
the Sao Paulo State University. 
The authors acknowledge funding from the Academy of Finland 
(FiDiPro scheme to RM) and the German Research Foundation  
(DFG Grant CH 707/5-1 to AGC).

\endpage

\endpage
\newpage

\renewcommand*{\citenumfont}[1]{A#1}
\renewcommand*{\bibnumfmt}[1]{[A#1]}
\numberwithin{figure}{section}
\renewcommand{\thefigure}{A\arabic{figure}}
\appendix\section{Supplementary Material}

In the Supplementary Material we include the video files illustrating the
change of polymer conformations for the two basic geometries, as
investigated in the main text. For each geometry, we fix the value of the
surface charge density $\sigma$ and the confinement dimensions $a$ and vary
the solution salinity. In doing so, at small $\kappa a$ values the
polyelectrolyte chain is rather in the adsorbed state, while for larger
$\kappa a$ the electrostatic polymer-surface screening gets stronger, the
system crosses the adsorption-desorption boundary, and the chain desorbs
from the interface (see also Fig. 6 of the main text). 
Parameters: the spherical cavity with the surface charge density of  
$\sigma=-0.1/(4 \pi)$ C/m$^2$, the sphere radius is $a=50$\AA, and the chain
polymerization degree is $N=100$, simulated at $\kappa a=1$ (video
1) and 10 (video 2). The cylindrical pore for the same values of 
$\sigma$, $N$, and $a$ simulated for $\kappa a=1$ (video 3) and 10 (video
4). Every video contains about 10$^6$ elementary simulation steps.
Note that the length of the trace shown is about 0.1 of the entire
trajectory length used in Fig. 6 to determine the
adsorption-desorption boundary.

\end{document}